\documentclass[twocolumn]{aastex631}

\usepackage{ae,aecompl}
\usepackage{subfigure}
\usepackage{graphicx}	% Including figure files
\usepackage{amsmath}	% Advanced maths commands
\usepackage{amssymb}	% Extra maths symbols
\usepackage{color,xcolor}
\usepackage{float}
\usepackage{multirow}
\usepackage{makecell}
\usepackage{xparse}
\usepackage{ulem}

\newcommand{\degree}{\hbox{$^\circ$}}

\newcommand{\msun}{$M_{\odot}$}
\newcommand{\lsun}{$L_{\odot}$}
\newcommand{\um}{$\mu$m}

\NewDocumentCommand{\fengwei}{ m g }{%
    \IfNoValueTF{#2}
        {\textcolor{teal}{[Fengwei: #1]}}
        {{\color{red}\sout{#1}} \textcolor{teal}{[Fengwei: #2]}}
}

\received{xxx}
\revised{xxx}
\accepted{xxx}

\shorttitle{The ALMA-QUARKS Survey:}
\shortauthors{Yang et al.}

\begin{document}
\title{\bf The ALMA-QUARKS survey: Evidence of a candidate high-mass prestellar core aside a bright-rimmed cloud IRAS 18290-0924}

\submitjournal{ApJ}

\author{Dongting Yang}
\affiliation{School of physics and astronomy, Yunnan University, Kunming, 650091, PR China}

\author{Hong-Li Liu}
\affiliation{School of physics and astronomy, Yunnan University, Kunming, 650091, PR China}
\correspondingauthor{Hong-Li Liu}
\email{hongliliu2012@gmail.com}

\author{Shengli Qin}
\affiliation{School of physics and astronomy, Yunnan University, Kunming, 650091, PR China}
\correspondingauthor{Shengli Qin}
\email{qin@ynu.edu.cn}

\author{Tie Liu}
\affiliation{Shanghai Astronomical Observatory, Chinese Academy of Sciences, 80 Nandan Road, Shanghai 200030, Peoples Republic of China}
\correspondingauthor{Tie Liu}
\email{liutie@shao.ac.cn}

\author{Anandmayee Tej}
\affiliation{Indian Institute of Space Science and Technology, Thiruvananthapuram 695 547, Kerala, India}
\correspondingauthor{Anandmayee Tej}
\email{tej@iist.ac.in}

\author{Siju Zhang}
\affiliation{Departamento de Astronom\'{i}a, Universidad de Chile, Las Condes, 7591245 Santiago, Chile}

\author{Xunchuan Liu}
\affiliation{Shanghai Astronomical Observatory, Chinese Academy of Sciences, 80 Nandan Road, Shanghai 200030, Peoples Republic of China}

\author[0000-0001-5950-1932]{Fengwei Xu}
\affiliation{Kavli Institute for Astronomy and Astrophysics, Peking University, 5 Yiheyuan Road, Haidian District, Beijing 100871, People's Republic of China}
\affiliation{Department of Astronomy, Peking University, 100871, Beijing, People's Republic of China}

\author{Guido Garay}
\affiliation{Departamento de Astronom\'ia, Universidad de Chile, Casilla 36-D, Santiago, Chile}
\affiliation{Chinese Academy of Sciences South America Center for Astronomy, National Astronomical Observatories, Chinese Academy of Sciences, Beijing, 100101, PR China}

\author{Lei Zhu}
\affiliation{Chinese Academy of Sciences South America Center for Astronomy, National Astronomical Observatories, Chinese Academy of Sciences, Beijing, 100101, PR China}

\author[0000-0002-7125-7685]{Patricio Sanhueza}
\affiliation{Department of Astronomy, School of Science, The University of Tokyo, 7-3-1 Hongo, Bunkyo, Tokyo 113-0033, Japan}

\author[0000-0002-0786-7307]{Xiaofeng Mai}
\affiliation{Shanghai Astronomical Observatory, Chinese Academy 
of Sciences, Shanghai 200030, People’s Republic of China}
\affiliation{School of Astronomy and Space Sciences, University of
Chinese Academy of Sciences, No. 19A Yuquan Road, Beijing 100049,
People’s Republic of China}

\author{Wenyu Jiao}
\affiliation{Shanghai Astronomical Observatory, Chinese Academy of Sciences, 80 Nandan Road, Shanghai 200030, Peoples Republic of China}

\author{Paul F. Goldsmith}
\affiliation{Jet Propulsion Laboratory, California Institute of Technology, 4800 Oak Grove Drive, Pasadena, CA 91109, USA}

\author{Sami Dib}
\affiliation{Max Planck Institute for Astronomy, Königstuhl 17, D-69117 Heidelberg, Germany}

\author[0000-0002-8586-6721]{Pablo Garc\'ia}
\affiliation{Chinese Academy of Sciences South America Center for Astronomy, National Astronomical Observatories, CAS, Beijing 100101, China}
\affiliation{Instituto de Astronom\'ia, Universidad Cat\'olica del Norte, Av. Angamos 0610, Antofagasta, Chile}

\author{Di Li}
\affiliation{New Cornerstone Science Laboratory, Department of Astronomy, Tsinghua University, Beijing 100084, China}

\author{Jinhua He}
\affiliation{Yunnan Observatories, Chinese Academy of Sciences, Kunming, 650216, Yunnan, PR China}
\affiliation{Chinese Academy of Sciences South America Center for Astronomy, National Astronomical Observatories, Chinese Academy of Sciences, Beijing, 100101, PR China}
\affiliation{Departamento de Astronom\'{i}a, Universidad de Chile, Las Condes, 7591245 Santiago, Chile}

\author{A.Y. Yang}
\affiliation{National Astronomical Observatories, Chinese Academy of Sciences, Beijing, 100101, PR China} 
\affiliation{Key Laboratory of Radio Astronomy and Technology, Chinese Academy of Sciences, A20 Datun Road, Chaoyang District, Beijing, 100101, PR China}

\author{Prasanta Gorai}
\affiliation{Rosseland Centre for Solar Physics, University of Oslo, PO Box 1029 Blindern, 0315 Oslo, Norway}
\affiliation{Institute of Theoretical Astrophysics, University of Oslo, PO Box 1029 Blindern, 0315 Oslo, Norway}

\author{Suinan Zhang}
\affiliation{Shanghai Astronomical Observatory, Chinese Academy of Sciences, 80 Nandan Road, Shanghai 200030, Peoples Republic of China}

\author{Yankun Zhang}
\affiliation{Shanghai Astronomical Observatory, Chinese Academy of Sciences, 80 Nandan Road, Shanghai 200030, Peoples Republic of China}

\author{Jianjun Zhou}
\affiliation{Xinjiang Astronomical Observatory, Chinese Academy of Sciences,
150 Science 1-Stree, Urumqi, Xinjiang 830011, People's Republic of China}

\author[0000-0002-5809-4834]{Mika Juvela}
\affiliation{Department of Physics, P.O. box 64, FI- 00014, University of Helsinki, Finland}

\author[0000-0002-9875-7436]{James O. Chibueze}
\affiliation{Department of Mathematical Sciences, University of South Africa, Cnr Christian de Wet Rd and Pioneer Avenue, Florida Park, 1709, Roodepoort, South Africa}
\affiliation{Department of Physics and Astronomy, Faculty of Physical Sciences, University of Nigeria, Carver Building, 1 University Road, Nsukka 410001, Nigeria}

\author{Chang Won Lee}
\affiliation{Korea Astronomy and Space Science Institute, 776 Daedeokdae-ro, Yuseong-gu, Daejeon 34055, Republic of Korea}
\affiliation{University of Science and Technology, Korea (UST), 217 Gajeong-ro, Yuseong-gu, Daejeon 34113, Republic of Korea}

\author{Jihye Hwang}
\affiliation{Korea Astronomy and Space Science Institute, 776 Daedeokdae-ro, Yuseong-gu, Daejeon 34055, Republic of Korea}

\author{Leonardo Bronfman}
\affiliation{Departamento de Astronom\'ia, Universidad de Chile, Casilla 36-D, Santiago, Chile}

\author{Xindi Tang}
\affiliation{Xinjiang Astronomical Observatory, Chinese Academy of Sciences,
150 Science 1-Stree, Urumqi, Xinjiang 830011, People's Republic of China}

\author{Archana Soam}
\affiliation{Indian Institute of Astrophysics, II Block, Koramangala, Bengaluru 560034, India}

\author{Tapas Baug}
\affiliation{S. N. Bose National Centre for Basic Sciences, Block-JD, Sector-III, Salt Lake City, Kolkata 700106, India}

\author{Yichen Zhang}
\affiliation{Department of Astronomy, Shanghai Jiao Tong University, 800 Dongchuan Rd., Minhang, Shanghai 200240, People's Republic of China }

\author{Swagat Ranjan Das}
\affiliation{Departamento de Astronom\'ia, Universidad de Chile, Casilla 36-D, Santiago, Chile}

\author{L. K. Dewangan}
\affiliation{Astronomy $\&$ Astrophysics Division, Physical Research Laboratory, Navrangpura, Ahmedabad 380009, India}

\author[0000-0002-5310-4212]{L. Viktor T\'oth}
\affiliation{Institute of Physics and Astronomy, E\"otv\"os Lor\`and University, P\'azm\'any P\'eter s\'et\'any 1/A, H-1117 Budapest, Hungary}
\affiliation{Faculty of Science and Technology, University of Debrecen, H-4032 Debrecen, Hungary}

\begin{abstract}
Although frequently reported in observations, the definitive confirmation of high-mass prestellar cores has remained elusive, presenting a persistent challenge in star formation studies. Using two-band observational data from the 3mm ATOMS and 1.3mm QUARKS surveys, we report a high-mass prestellar core candidate, C2, located on the side of the bright-rimmed cloud IRAS 18290-0924. The C2 core identified from the 3mm continuum data of the ATOMS survey ($\sim$2\,\arcsec, $\rm\sim 10000\,au$ at 5.3\,kpc) has a mass ranging from 27-68\,\msun\, for temperatures 10-22\,K within a radius of $\sim$2800\,au. The highest-resolution ($\sim$0.3\arcsec, $\rm\sim 1500\,au$) observations of this source presented to date from the QUARKS survey reveal no evidence of further fragmentation.
Further analysis of a total $\sim$10\,GHz band width of molecular line survey does not find star-formation activity (e.g., outflows, ionized gas) associated with the core, with a few molecular lines of cold gas detected only. Additionally, virial analysis indicates the C2 core is gravitationally bound ($\alpha_{\rm vir} \sim0.1-0.3$) and thus could be undergoing collapse toward star formation. 
These results strongly establish a candidate for a high-mass prestellar core, contributing to the very limited number of such sources known to date.
\end{abstract}


\section{Introduction} \label{sec:intro}
Understanding star formation requires tracing the evolution of molecular gas from its earliest prestellar phase to the formation of protostars \citep{2000prpl.conf...59A}. For high-mass stars, which typically form in clusters, their far distances, short evolutionary timescales, and intense stellar feedback make it particularly challenging to directly observe the formation process \citep{2003ARA&A..41...57L,2009ApJ...696..268Z}.
Despite extensive observational and theoretical research over the past few decades, the exact mechanisms driving high-mass star formation remain uncertain \citep{2018ARA&A..56...41M}.

Currently, the two widely accepted models of high-mass star formation are the ``turbulent core accretion" model \citep{2003ApJ...585..850M} and the ``competitive accretion" model \citep{2001MNRAS.323..785B}.
The former posits the existence of a pre-assembled massive gas reservoir within a clump, which contains the final mass required to form a high-mass star. This structure, referred to as a massive prestellar core, maintains stability against collapse  due to support mechanisms that include turbulence and magnetic fields. In contrast, in the ``competitive accretion" model, the massive gas clumps initially fragment into thermal-Jeans-like low-mass cores. Subsequently, these cores competitively accrete gas from the clump-scale reservoir. Here, the more massive cores located deeper in the gravitational potential well of the clump exhibit larger accretion efficiency and accelerated mass growth.
Furthermore, accretion flows from filamentary structures associated with
hub-filament systems enable additional gas mass flow into the cores from larger scales (>1\,pc), facilitating the more efficient and rapid formation of massive stars in the central hubs (e.g., \citealt{2013A&A...555A.112P,2019MNRAS.490.3061V,2020ApJ...900...82P,2023ApJ...953...40Y,2024MNRAS.534.3832D}).

Extensive observational investigations from SMA, NOEMA and ALMA revealed that massive gas clumps predominantly fragment into low-mass cores, which subsequently accrete gas and grow mass to form massive stars \citep{2015ApJ...804..141Z,2019ApJ...886...36S,2018A&A...617A.100B,2019ApJ...886..102S,2021A&A...646A..25Z,2024ApJ...966..171M,2025arXiv250305663C}. 
 In comparison, case studies as well as large-sample statistical analysis have not found conclusive evidence for the existence of massive prestellar cores (e.g., \citealt{2007A&A...476.1243M,2010A&A...524A..18B,2019A&A...622A..99L,2013ApJ...779...96T,2023ApJ...945...81J,2023ApJ...950..148M}). 

Based on numerous observational results, \cite{2018ARA&A..56...41M} proposed an evolutionary scenario for massive star formation that excludes such high-mass prestellar cores. However, the detection of a limited number of massive prestellar core candidates ($\gtrsim  16$\msun\,, e.g., \citealt{2013A&A...558A.125D,2014ApJ...796L...2C,2014MNRAS.439.3275W,2023A&A...675A..53B,2024ApJ...961L..35M,2024ApJS..270....9X,2025A&A...696A..11V}) continues to sustain the debate about massive star formation mechanisms.
The detection of specific molecular lines distinguishes the prestellar or protostellar nature of dense cores (e.g., \citealt{2019ApJ...886..102S,2025ApJS..280...33Y}), as star formation activity within protostellar cores increases the ambient temperature, thus releasing complex organic molecules (COMs, e.g., $\rm CH_3OH$) from grain mantles into the gas phase, which then become observable \citep{2014A&A...563A..97G}.
Reliable identification of high-mass prestellar cores is challenging without a  broad frequency range survey of molecular lines (e.g., \citealt{2018A&A...618L...5N,2019A&A...626A.132M,2013A&A...558A.125D,2014MNRAS.439.3275W,2025A&A...696A..11V}).
In addition, their interpretation is further complicated by the ambiguous absence of outflows and the observed presence of fragmentation
\citep{2013A&A...558A.125D,2024ApJ...961L..35M}.

 \begin{figure}[ht!]
    \centering
    \includegraphics[angle=0, width=0.45\textwidth]{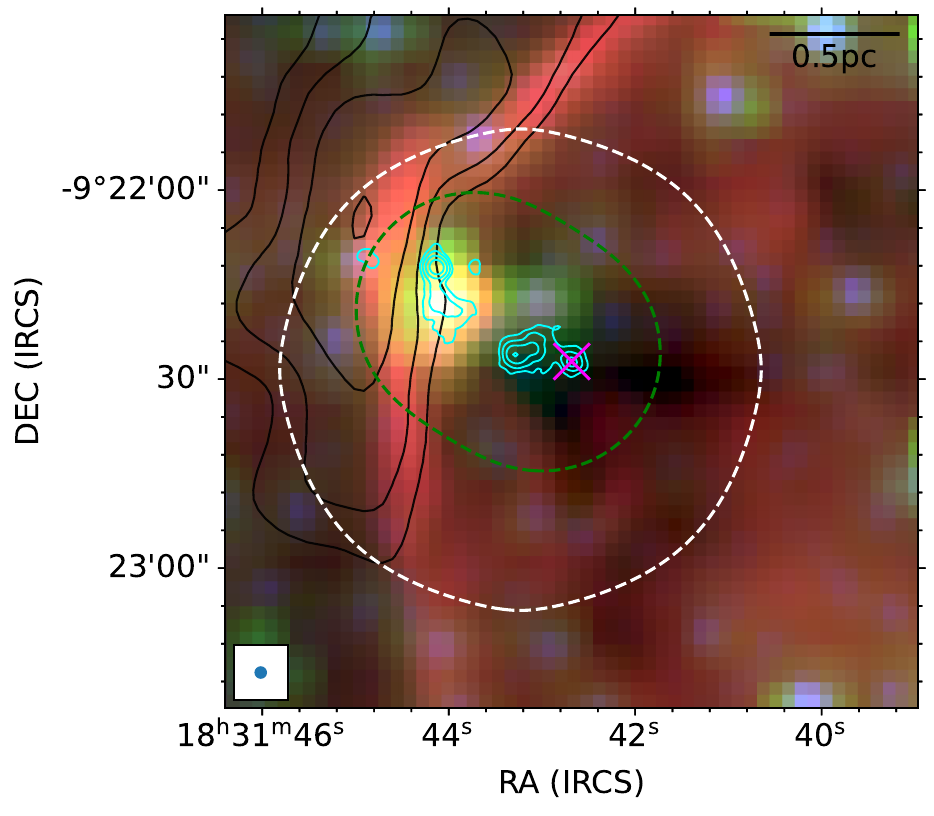} 
    \caption{Three-colour image of Spitzer 8\,\um\ (red), 4.5\,\um\ (green), and 3.6\,\um\ (blue) of the region associated with I18290.
    %region. 
    The black contours represent the 1.28\,GHz MeerKAT emission. The white and green dashed circles show the FoV of the ATOMS and QUARKS survey, respectively. ATOMS 3\,mm dust continuum emission is shown as the cyan contours, with levels starting at 3\,$\sigma_{\rm 3\,mm}$ ($\rm\sim 0.1mJy~beam^{-1}$) and increasing as $[6, 12, 24, 48]\times \sigma_{\rm 3\,mm}$. 
    The purple cross symbol shows the location of Core\,2 \citep{2023MNRAS.520..322Z}.
    The synthesized beam size of the ATOMS survey and the 0.5\,pc scale bar are shown in the lower left corner and upper right corner, respectively. } 
    \label{fig:RGB}    
\end{figure}

The search for high-mass prestellar cores has primarily focused on infrared-dark environments, as only these regions are believed to harbour massive cloud cores in their earliest evolutionary stages
(e.g., \citealt{2013ApJ...779...96T,2017ApJ...841...97S,2018ARA&A..56...41M}). However, recent studies suggest that massive prestellar cores may also exist within infrared-bright environments (e.g., \citealt{2019ApJ...886..102S,2024ApJS..270....9X}). 
This is likely due to elevated temperatures and heightened turbulence in these regions. The increased temperature raises the Jeans mass, which in turn facilitates the formation of high-mass prestellar cores—particularly when combined with the additional support provided by magnetic fields.

Bright-rimmed clouds (BRCs) constitute a distinct class of infrared-bright star-forming environments \citep{1991ApJS...77...59S,1994ApJS...92..163S}. 
They are characterized by an ionized surface on one side, created by intense ultraviolet radiation from nearby OB stars, which forms an over-pressured ionized boundary layer.
This boundary layer can be traced via free-free emission or radio recombination lines. 
Investigations of massive star formation in BRCs, particularly those focusing on embedded high-mass prestellar cores, remain scarce. Thus, identifying such cores within BRCs is crucial for constraining models of massive star formation.

The target of this study is a massive star formation region associated with the BRC, IRAS 18290–0924 (hereafter I18290, see Figure\,\ref{fig:RGB})
with estimated clump mass $\sim$1500\,\msun\ and luminosity $\sim10^4$\,\lsun\,\citep{2018MNRAS.473.1059U,2020MNRAS.496.2790L}.
Located at a kinematic distance of $\sim5.34\pm0.5$\,kpc \citep{2018MNRAS.473.1059U,2021A&A...646A..74M,2014ApJ...790...84L}, this region hosts dense cores that display a small-scale (<1\,pc) age sequence aligned with the direction of ionization, indicative of subsequent star formation triggered by radiation-driven implosion \citep{2023MNRAS.520..322Z}. 
Among them, the C2 core, situated farthest from the bright rim, has been proposed as a high-mass prestellar core candidate
\citep{2023MNRAS.520..322Z}.
In this work, we utilize ALMA two-band observations (see Sect.\,\ref{sec:obs}) to perform a comprehensive analysis of the C2 core with the aim of assessing its evolutionary stage and evaluating its candidacy as a high-mass prestellar core.

\section{Observations} \label{sec:obs}
The I18290 region has been observed both in the ALMA Three-millimeter Observations of Massive Star-forming regions (ATOMS, Project ID: 2019.1.00685.S; \citealt{2020MNRAS.496.2790L,2021MNRAS.505.2801L}) survey and the Querying Underlying mechanisms of massive star formation with ALMA-Resolved gas Kinematics and Structures (QUARKS, Project ID: 2021.1.00095.S; \citealt{2024RAA....24b5009L,2024RAA....24f5011X,2025ApJS..280...33Y}) survey. 

Observations for the ATOMS survey were conducted using ALMA 7-m Atacama Compact array (ACA) and the 12-m arrays at Band\,3 ($\rm\sim 3\,mm$), targeting 146 massive star-forming protocluster clumps. Further details regarding the ATOMS survey are available in \cite{2020MNRAS.496.2790L}. ACA and 12-m array data from the ATOMS survey were combined to produce continuum and images and line cubes, resulting in an angular resolution $\sim2$\,\arcsec\ for continuum emission. 
For I18290, the combined continuum data have rms noise ($\sigma_{\rm 3\,mm}$) $\sim0.1\,\rm mJy~beam^{-1}$. For line cubes of the ATOMS survey, eight spectral windows (SPWs) were configured, including six high-spectral-resolution SPWs ($\sim\rm 0.2-0.4~km~s^{-1}$, e.g., $\rm H^{13}CO^+\,(1-0)$) and two wide SPWs ($\sim\rm 1.8\,GHz$) with spectral-resolution of $\sim\rm 1.6~km~s^{-1}$.

 \begin{figure*}[ht!]
    \centering
    \includegraphics[angle=0, width=1.\textwidth]{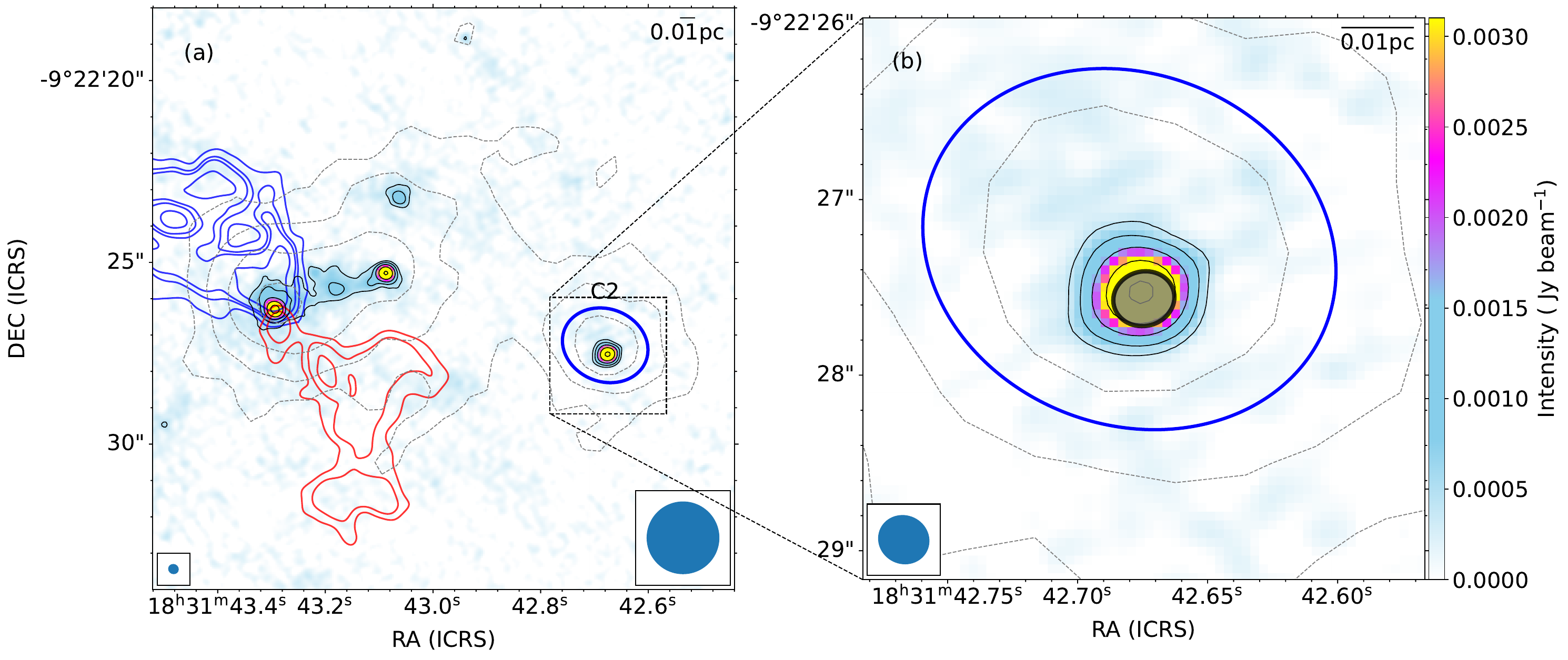} 
    \caption{QUARKS 1.3\,mm dust continuum image (color scale) overlaid with ATOMS 3\,mm dust continuum emission (in gray dashed contours, identical to those in Figure\,\ref{fig:RGB}).
    The black contours are at levels of [3, 6, 12, 24, 48]\,$\times \sigma_{\rm 1.3\,mm}$, with $\sigma_{\rm 1.3\,mm}\sim0.14\rm mJy~beam^{-1}$. The blue ellipses correspond to the FWHM sizes of the C2 core estimated from the ATOMS 3\,mm image using the task {\it CASA imfit}. The synthesized beam sizes of the QUARKS and ATOMS survey are shown in the left and right bottom (panel\,a), respectively. The 0.01\,pc scale bar is shown in the upper right corner. Panel\,(a): CO (2--1) outflow is overlaid on the QUARKS 1.3\,mm dust continuum map. The blue and red contour levels are [3, 6, 12, 24, 48]\,$\times \sigma_{\rm co}$, with $ \sigma_{\rm co} \sim\rm 0.08~Jy~beam^{-1}~km~s^{-1}$ for the blue lobes and $\sim\rm 0.06~Jy~beam^{-1}~km~s^{-1}$ for the red lobes. The corresponding velocity ranges are plus/minus $\rm ~5-30~km~s^{-1}$ relative to $\rm V_{LSR}$ ($\sim\rm 84.5~km~s^{-1}$) of the C2 core. Panel\,(b): zoom-in view of the C2 core. The black ellipses correspond to the FWHM sizes of the condensation estimated from the QUARKS 1.3\,mm image using the task {\it casa imfit}.} 
    \label{fig:cont}    
\end{figure*}

\begin{deluxetable*}{ccccccccccc}\label{table:casaimfit}
\tabletypesize{\scriptsize}
% \tablewidth{0pt} 
\tablecaption{Observed parameters of core and condensation from {\it casa~imfit}.} 
% \tablecomments{}
% \centering
\tablehead{
\colhead{Source} & \colhead{R.A.} & \colhead{Dec.}   & \colhead{$\rm FWHM_{maj}$} & \colhead{$\rm FWHM_{min}$} &\colhead{PA} & \colhead{$\rm FWHM_{maj}^{Decon}$} & \colhead{$\rm FWHM_{min}^{Decon}$}  & \colhead{$F_{\nu}^{\rm int}$} & \colhead{$F_{\nu}^{\rm peak}$}&\colhead{Data}\\
\colhead{}& \colhead{(ICRS)}& \colhead{(ICRS)}&  \colhead{($\arcsec$)} & \colhead{($\arcsec$)} & \colhead{($\degree$)} & \colhead{($\arcsec$)} & \colhead{($\arcsec$)} & \colhead{($\rm mJy$)} & \colhead{($\rm mJy\,beam^{-1}$)} \\ }
\colnumbers 
\startdata 
 Core& 18:31:42.68 & -09:22:27.28  & 2.4$\pm$0.1 & 2.0$\pm$0.1 & 68.8$\pm$8.8& 1.2$\pm$0.3 & 0.9$\pm$0.3 &2.15$\pm$0.13 & 1.66$\pm$0.06&Band3\\
% \hline
 Condensation& 18:31:42.67 & -09:22:27.53  & 0.35$\pm$0.01 & 0.31$\pm$0.01 & 103.9$\pm$3.8 & 0.21$\pm$0.01 & 0.12$\pm$0.01 & 11.21$\pm$0.22 & 8.35$\pm$0.11&Band6\\
\enddata
\begin{flushleft}
\end{flushleft}

\end{deluxetable*}

The QUARKS survey is a follow-up to the ATOMS survey, which observed 139 massive star-forming protocluster clumps through 156 single-pointings.
QUARKS observations were obtained with three different ALMA configurations at Band\,6 ($\sim\rm1.3\,mm$), including relatively low ($\sim$5\arcsec), moderate ($\sim$1\arcsec) and high ($\sim$0.3\arcsec) angular resolution, which were performed using the ACA 7-m array, the ALMA 12-m compact array C-2 (TM2) and extended C-5 (TM1) configurations, respectively. More details regarding the QUARKS survey are available in \cite{2024RAA....24b5009L}, \cite{2024RAA....24f5011X}, and \cite{2025ApJS..280...33Y}. Combining three configuration observations from the QUARKS survey for both continuum and line data yielded a synthesized beam size of $\sim0.3$\,\arcsec. The combined QUARKS continuum data of the I18290 region have a typical sensitivity of rms ($\sigma_{\rm 1.3\,mm}$) $\sim0.14\,\rm mJy~beam^{-1}$. Four SPWs were configured for the QUARKS survey, with a bandwidth of $\sim$1.8\,GHz and a velocity resolution of $\sim\rm 1.3\,km\,s^{-1}$ for each SPW.

\begin{deluxetable*}{cccccccc}\label{table:calculated}
% \tabletypesize{\scriptsize}
% \tablewidth{0pt} 
\tablecaption{Physical parameters of the core and condensation.} 
% \tablecomments{}
% \centering
\tablehead{
\colhead{Source} & \colhead{Radius$^a$}  & \colhead{$\sigma_{\rm obs}$} & \colhead{Temperature}   & \colhead{Mass} & \colhead{$n_{\rm H_2}$}& \colhead{$\alpha_{\rm vir}$} &\colhead{Data} \\
\colhead{G022.3501+00.0697}& \colhead{(au)}& \colhead{($\rm km~s^{-1}$)} & \colhead{(K)}&  \colhead{(\msun)} & \colhead{($\rm 10^8~cm^{-3}$)}  &  \\ }
\colnumbers 
\startdata 
Core & 2800$\pm$500 & 0.62$\pm$0.37 &10  & 68.2$\pm$13.4 & 0.9$\pm$0.5  &0.10$\pm$0.03 & Band\,3\\
     &  &  &   22 & 26.9$\pm$5.2 & 0.4$\pm$0.2 &0.27$\pm$0.09 \\
\hline
Condensation& 400$\pm$50 & 0.50$\pm$0.20& 10  & 19.1$\pm$3.6 & 79.0$\pm$20.0 &  0.03$\pm$0.01 & Band\,6\\
& && 22  & 6.1$\pm$1.2 & 25.5$\pm$6.5 &  0.12$\pm$0.04 \\
\enddata
\begin{flushleft}
$^a$ Beam-dencovolved.
\end{flushleft}

\end{deluxetable*}

\section{Results}\label{sec:res}
\subsection{Parameter estimation of core and condesation
}
% C2 core has been classified as a '$unknown$' source in \cite{2021MNRAS.505.2801L}. 
To ensure consistent parameter estimation, the {\it imfit} task in {\it CASA} \citep{2022PASP..134k4501C} was used to extract structures from both ATOMS 3\,mm and QUARKS 1.3\,mm dust continuum maps.
The {\it imfit} task involves a two-dimensional Gaussian fit to the continuum emission of structures.
We adopt a hierarchical terminology by designating ATOMS-identified structures as cores and QUARKS-resolved substructures as condensations (see Figure\,\ref{fig:cont}). Note that these terms represent manifestations of the same structure at different spatial resolutions. Only a single compact condensation is observed in the core C2 (see Figure\,\ref{fig:cont}\,(b), indicating that the cloud core
has not undergone fragmentation at the high angular resolution ($\sim$0.3\,\arcsec) of the QUARKS survey. The observed parameters of the core and the condensation retrieved with the {\it imfit} task are listed in Table\,\ref{table:casaimfit}.

If we assume that the 3\,mm and 1.3\,mm continuum emission is optically thin and mainly arises from thermal dust radiation, the masses of the core and condensation can be estimated using the following expression:

\begin{equation}
    M_{\rm gas}=\frac{R_{\rm gd}S^{\rm int}_\nu D^2}{\kappa_{\nu}B_{\nu}(T_{\rm dust})},
\end{equation}
where $R_{\rm gd}$ is the ratio of gas to dust (assumed to be 100), $S^{\rm int}_{\nu}$ is the integrated flux derived and $D$ is the distance to the source.
The dust opacity, $\kappa_{\nu}$, is taken as 0.18\,$\rm cm^2~g^{-1}$ and 0.9\,$\rm cm^2~g^{-1}$ for the 3\,mm and 1.3\,mm continuum emission, respectively \citep{1994A&A...291..943O}. $B_{\nu}(T_{\rm dust})$ is the Planck function at a given dust temperature. 

Dust temperatures as low as 10–15\,K have been reported in several IRDC environments harbouring starless cores without internal protostellar heating \citep{2013ApJ...773..123S,2023A&A...675A..53B,2024ApJ...961L..35M}. 
In this case, given the starless nature of the C2 core (see Sect.\,\ref{subsec:3.2}), we adopt 10\,K as the lower limit for the dust temperature. Further, the average dust temperature of the natal clump (22\,K, derived from the spectral energy distribution, \citealt{2018MNRAS.473.1059U}) is considered as the upper limit. For these temperature limits the estimated masses of the C2 core and condensation are 26.9-68.2\,\msun\ and 6.1-19.1\,\msun\ at 3\,mm and 1.3\,mm, respectively.

In addition, assuming a spherical geometry, the average number density can be calculated as $n_{\rm H_2} = \frac{3M_{\rm core}}{4\pi R_{\rm c}^3 \mu m_{\rm H}}$, where $\mu = 2.8$ is the mean molecular weight of the hydrogen molecule and $m_{\rm H}$ is the mass of the hydrogen atom \citep{2008A&A...487..993K}. The effective radius, $R_{\rm c}$, is estimated as $\sqrt{\rm FWHM^{Decon}_{maj}\times FWHM^{Decon}_{min}}/2\times D$, where $\rm FWHM^{Decon}_{maj}$ and $\rm FWHM^{Decon}_{min}$ are the deconvolved FWHM of the major and minor axes, respectively (see Col.\,7-8 in Table\,\ref{table:casaimfit}). The radii of the C2 core and condensation are $\sim 2800\pm500$\,au and $\sim 400\pm20$\,au, respectively. The average number densities are $\rm 0.4-1.0\times 10^8~cm^{-3}$ and $\rm 2.6-7.9\times 10^9~cm^{-3}$, respectively. The estimated physical parameters of the core and condensation are listed in Table\,\ref{table:calculated}.

 \begin{figure*}[ht!]
    \centering
    \includegraphics[angle=0, width=1.0\textwidth]{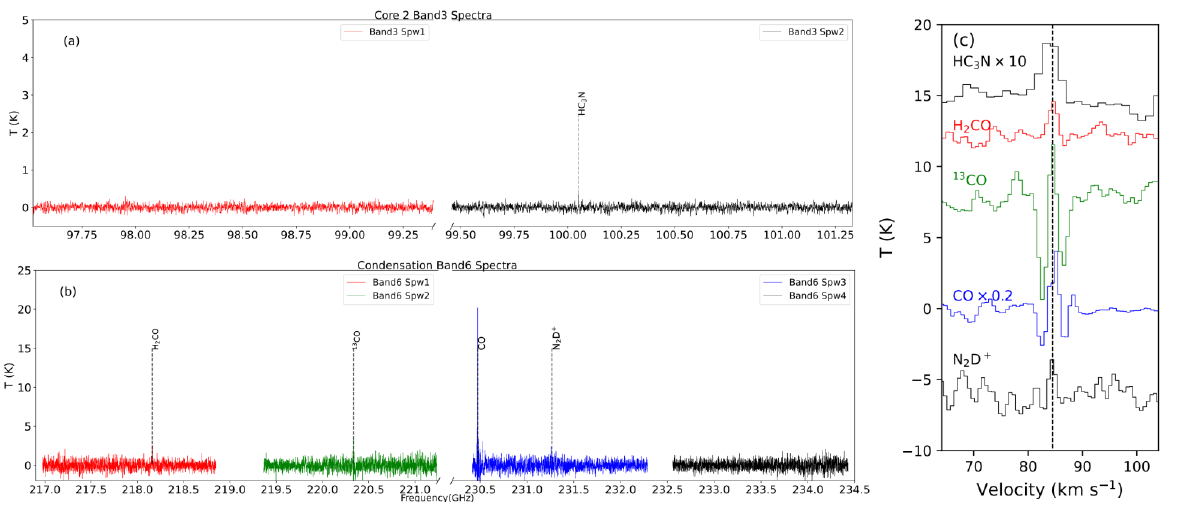} 
    % \caption{Average spectra of C2 core and condensation from the ATOMS survey (panel\,a) and from the QUARKS survey (panel\,b).} 
    \caption{Average spectra of the C2 core/condensation. (a): average spectra extracted from two ATOMS wideband SPWs over the C2 core. (b): Same as panel\,a but for 
    four QUARKS SPWs over the C2 condensation. (c): zoom-in view of the molecular lines detected in the six SPWs (corresponding to panels\,a and b). The dashed line indicates the systemic velocity of the C2 core ($\rm\sim 84.5\,km\, s^{-1}$). Note that the absorption dips present in both $^{13}$CO and CO spectra are artifacts arising from the missing flux by ALMA, which will be addressed for future more in-depth studies by combining our QUARKS data with new single-dish observations.
    }
    \label{fig:spectrum}    
\end{figure*}

\subsection{Evolutionary stage of the C2 Core}\label{subsec:3.2}
Starless cores lack any star formation signatures, such as outflows traced by CO emission (e.g., \citealt{2018ApJS..235....3Y,2019ApJ...886..102S,2022A&A...658A.160Y,2022MNRAS.510.3389U}). 
We present the averaged spectra obtained from the ATOMS (two SPWs) and QUARKS (four SPWs) surveys for the C2 core and condensation in Figure \ref{fig:spectrum}\,(a) and Figure \ref{fig:spectrum}\,(b), respectively. Molecular line emission of $\rm HC_3N$ is detected in the ATOMS Spw2 spectrum. The QUARKS spectra exhibits $\rm ^{13}CO$ (2-1), CO (2–1), and H$_2$CO~(3$_{0,3}$–2$_{0,2}$) molecular lines. Additionally, $\rm N_2D^+$ (3-2) emission is marginal detected with a low signal-to-noise ratio,
which can be confirmed in the zoom-in image of the detected spectral lines in Figure\,\ref{fig:spectrum}\,c (see more in Sect.\,\ref{subsect:3.3}). 

\cite{2023MNRAS.520..322Z} used SiO and $\rm HCO^+$ emission from the ATOMS survey to search for outflows in the I18290 region, concluding that no outflows were detected toward the massive core C2. 
To confirm this result, we analyze the CO (2–1) emission from the QUARKS survey, which provides enhanced resolution to investigate potential outflows in this region.
The CO line profile is integrated in a velocity range of $\rm 5-30~km~s^{-1}$ relative to the systemic velocity ($\rm\sim 84.5\,km~s^{-1}$), and their spatial distribution is shown by the blue and red contours in the Figure\,\ref{fig:cont}\,(a). 
No CO outflow is detected towards the massive core C2 in the I18290 region. In a recent study based on ATOMS data, \cite{2025ApJ...987..197H} proposed the $\rm HC_3N$ transition as an effective tracer of low-velocity components of outflows. They reported the detection of an outflow associated with the I18290 region. A careful inspection of Figure\,A1 in \cite{2025ApJ...987..197H} confirms that the identified outflow is not associated with the C2 core. Furthermore, no line wings were identified in the H$_2$CO~(3$_{0,3}$–2$_{0,2})$ transition, which is also considered as an outflow tracer \citep{2024RAA....24b5009L}. The absence of the outflow signature in the line transitions investigated is
consistent with the findings of \cite{2023MNRAS.520..322Z}. 
In addition, weak detection of the $\rm N_2D^+$ line toward the C2 core indicates the presence of cold dense gas \citep{2005ApJ...619..379C, 2024RAA....24b5009L}. Due to the spatial filtering-out effect of the interferometric observations, it is likely that that detected $\rm N_2D^+$ line emission probes the C2 core rather than the extended envelope. 
Therefore, this evidence of the cold dense gas, coupled with the absence of detected outflows, supports the starless nature of the C2 core.

\subsection{Dynamical stability  of C2 Core and condensation}\label{subsect:3.3}
Understanding the stability of the unfragmented C2 core against gravitational collapse is essential to assess the potential formation of a high-mass star in this core. This can be assessed by estimating the virial parameters ($\alpha_{\rm vir}$), which, for an ideal spherical structure of uniform density, is given by \citep{1992ApJ...395..140B,2007ApJ...661..262D}:

\begin{equation}
    \alpha_{\rm vir}=\frac{5\sigma_{\rm eff}^2R_{c}}{\rm G{\it M_{\rm c}}}\,,
\end{equation}
where $\sigma_{\rm eff}$ is the effective sound speed.  $R_{\rm c}$ and $M_{\rm c}$ are the radius and mass of the structure, respectively. $\sigma_{\rm eff}$ can be substituted by the total velocity dispersion ($\sigma_{\rm tot}^2=\sigma_{\rm th}^2+\sigma_{\rm nt}^2$). The thermal velocity dispersion is $\sigma_{\rm th}=\sqrt{k_BT/\mu_p m_{H}}$, where $\mu_{p}=2.33$ is the mean molecular weight per free particle \citep{2008A&A...487..993K}.
In this study, the molecular line emission of $\rm H^{13}CO^+\,(1-0)$ from the ATOMS survey and $\rm N_2D^+\,(3-2)$ from the QUARKS survey were used to estimate the contribution of nonthermal motion from turbulence to the C2 core and condensation, respectively. 
The ATOMS $\rm H^{13}CO^+\,(1-0)$ line emission provides high velocity resolution ($\rm \sim 0.2\,km\,s^{-1}$) and is a probe typical of dense core envelopes (e.g., $n_{\rm crit}\sim10^5\,\rm cm^{-3}$, \citealt{2014A&A...563A..97G}). In contrast, for the higher angular resolution QUARKS data, $\rm N_2D^+\,(3-2)$ line emission was adopted to serve as kinematic tracers for colder and denser conditions \citep{2005ApJ...619..379C,2007A&A...470..221C}.
Despite the low signal-to-noise ratio, marginal detection of $\rm N_2D^+$ emission is confirmed by cross-checking with other detected molecular lines of the C2 core (see Figure\,\ref{fig:spectrum}).
The nonthermal velocity dispersion is $\sigma_{\rm nt}=(\sigma_{\rm line}^2-\frac{k_BT}{m_{\rm line}})^{1/2}$, where $k_B$ is the Boltzmann constant and $m_{\rm line}$ is the molecular mass of the observed molecule ($m_{\rm line}=30m_{H}$ for the $\rm H^{13}CO^+$ line and $m_{\rm line}=32m_{H}$ for $\rm N_2D^+$). The observed dispersions ($\sigma_{\rm obs}$) are $\sim$0.62\,$\rm km~s^{-1}$ for $\rm H^{13}CO^+$ of the C2 core and $\sim$0.50\,$\rm km~s^{-1}$ for $\rm N_2D^+$ of condensation, as shown in Figure\,\ref{fig:h13co}.

The total velocity dispersion is estimated to be $\sim$0.65\,$\rm km~s^{-1}$ for $\rm H^{13}CO^+$ and $\sim$0.53\,$\rm km~s^{-1}$ for $\rm N_2D^+$. From these values, we calculate $\alpha_{\rm vir}$ to be $\sim$0.1–0.3 for both the C2 core and the condensation, in the temperature range of 10–22\,K. Following Eq.\,5 of \citealt{2011A&A...530A.118P} with the magnetic field contribution is considered, 
% the magnetic virial parameter is as follows:
% \begin{equation}\label{eq:mg}
%     \alpha_{B\,,\rm vir}=\frac{5R_{c}}{\rm G{\it M_{\rm c}}}(\sigma_{\rm eff}^2+\frac{v^2_A}{6})\,,
% \end{equation}}
% where the Alfv\`en velocity $v_A=B/(\mu_0\rho)^{1/2}$, and $\mu_0$ is the permeability of the free space.
such low values of $\alpha_{\rm vir} << 1$ are possible in the case of magnetized cloud fragments provided the magnetic field is strong $\sim1\,$mG \citep{2013ApJ...779..185K}. 
Recent observational measurements using the Davis-Chandrasekhar-Fermi method do yield field strengths of 1-10\,mG in massive star-forming regions (e.g., \citealt{2021ApJ...923..204C,2021ApJ...913...85H,2022ApJ...941...51H,2024ApJ...972L...6S,2024ApJ...974..257Z,2025ApJ...980...87S,2025ApJ...985..222H}), and smaller values of $\sim 0.1-1\,$mG in starless/prestellar core \citep{2006MNRAS.369.1445K,2020ApJ...900..181K,2021ApJ...907...88P,2023AJ....165..198H}. 

The stability analysis is consistent with the observed fragmentation scenario of the C2 core, where only a single smaller-scale condensation is detected. Several theoretical and numerical studies suggest that magnetic fields, by providing support against gravity, play a key role in suppressing fragmentation. In a recent study, \cite{2021ApJ...912..159P} studied 18 massive dense cores to investigate this correlation. Though not robustly observed, these authors discuss the possible influence of the magnetic field on the fragmentation process. Based on the above arguments, one can infer C2 core to be an unfragmented, massive, and subvirial entity in a strong magnetized environment. This core is therefore gravitationally bound and has the potential to collapse toward for star formation.

\section{Discussion and conclusions}
Despite the fact that \cite{2018ARA&A..56...41M} have proposed an evolutionary scheme for high-mass stars that excludes the massive prestellar phase, the empirical identification of several high-mass prestellar cores provides strong evidence for their existence. There are several examples from previous literature.
\cite{2013A&A...558A.125D} reported the detection of a high-mass prestellar core candidate (CygX-N53-MM2) with mass of $\sim$21\,\msun\ and radius of 2500\,au, which may be however associated with tentative outflows.
\cite{2014MNRAS.439.3275W} have also observed a candidate (i.e.; G11.11-P6-SMA1, mass of $\sim$28\,\msun, radius of $\sim$2000\,au) without any signatures of CO outflows in SMA observations. However, this source has yet to be confirmed by higher-angular-resolution ALMA observations. 
In the ALMA-IMF survey, \cite{2025A&A...696A..11V} identified 12 prestellar cores of masses greater than 16\,\msun, four of which are above 30\,\msun. These sources have no associated CO and/or SiO outflows detected. 
However, a comprehensive survey of molecular lines with sufficiently broad bandwidths has not been conducted for these massive candidates, which is vital for accurately classifying prestellar cores. For instance, detection of COMs, such as $\rm CH_3OH$, in the 4.8\,GHz SPWs of the high-mass prestellar core candidate W43-MM1, typically linked with hot cores, raises questions about its classification. This suggests that W43-MM1 is more likely in the early protostellar phase rather than a high-mass prestellar core candidate.\citep{2018A&A...618L...5N,2019A&A...626A.132M}.
In addition, albeit with an examination of a wide-ranging molecular line survey, the classification of several other massive prestellar core candidates remains uncertain. This includes C2c1a \citep{2023A&A...675A..53B} and MM1-C and MM1-E1 \citep{2024ApJ...961L..35M}. For example, C2c1a is linked with faint CO outflows, as depicted in Figure 3 of \citealt{2023A&A...675A..53B}, suggesting the beginning of protostellar activity. Additionally, high-resolution imaging of MM1-C indicates fragmentation \citep{2024ApJ...961L..35M}, questioning its classification as a singular high-mass prestellar core.

In this study, we present the C2 core in the I18290 massive star-forming region as a prestellar core candidate, based on ALMA two-band observations from the ATOMS and QUARKS surveys.
Under the assumption of a 10\,K dust temperature, we estimate the physical parameters of the C2 core using 3\,mm dust continuum emission from the ATOMS survey. 
The mass, radius, and average number density of the core were estimated to be $\sim$27-68\,\msun, $\sim$2800\,au, and $\sim\rm 10^8\,cm^{-3}$, respectively.
In striking contrast to the clustered, centrally located environment of the 12 massive prestellar cores reported by \cite{2025A&A...696A..11V}, it is worth noting that the C2 core is rather isolated. 

Subsequent higher-angular-resolution ($\sim 0.3$\,\arcsec) 1.3\,mm dust continuum data from the QUARKS survey revealed that the C2 core remains unfragmented, hosting only a central compact condensation
%structure 
with a mass of $\sim$19\,\msun, a radius of $\sim$400\,au, and average number density of $\sim\rm 10^9\,cm^{-3}$. This condensation exhibits highly compact dust emission with nearly circular symmetry (ellipticity $\sim 1.1$) 
(see Figure\,\ref{fig:cont}). Including the results from \citet{2023MNRAS.520..322Z}, no outflows (tracers, $\rm HCO^+$, SiO, and CO)  were detected nor any YSO was identified (see Sect.\,4.2 in \citealt{2023MNRAS.520..322Z}) with the massive C2 core. 
Furthermore, we only detected six molecular line emissions within six SPWs (two from the ATOMS survey, $\sim 4$\,GHz; four from the QUARKS survey, $\sim8$\,GHz; see Figure\,\ref{fig:spectrum}). None of the detected molecular line transitions are tracers of dense warm gas.
Our stability analysis reveals that thermal and turbulent support alone is insufficient to counteract gravitational collapse, giving a virial parameter of $\sim 0.1-0.3$. These low values of the virial parameter suggest a strong magnetized environment which lends further support to the unfragmented nature of the C2 Core.

The identification of the C2 core in I18290 presents a case study in support of the turbulent core model. The influence of environment leading to core mass growth can be ruled out-- the more evolved the natal environment, the more massive the embedded core. Though associated with the BRC I18290, the C2 core is located in an IR-dark lane (see Figure\,\ref{fig:RGB}) strongly indicating its very early evolutionary stage and its massive nature from the beginning. 
Additionally, the isolated nature of the C2 prestellar candidate provide a template for more in-depth studies, such as modeling and chemical evolution analysis.
 Although our study has outlined its fundamental physical properties,
a conclusive classification also hinges on the chemical evolution analysis. Our upcoming investigations aim to determine the chemical nature of this prestellar core candidate through quantitative analysis of molecular abundances and comparisons with chemical models, with an emphasis on the elevated abundances of deuterated isotopologues (e.g., \citealt{1999ApJ...523L.165C,2022ApJ...929...13C,2005ApJ...619..379C}).

\section*{Acknowledgements}
This work has been supported by the National Key R\&D Program of China (No.\,2022YFA1603101), the National Natural Science Foundation of China (NSFC) through grants No.12073061, No.12122307, and No.12033005.
H.-L. Liu is supported by Yunnan Fundamental Research Project (grant No.\,202301AT070118, 202401AS070121), and by Xingdian Talent Support Plan--Youth Project. 
D.-T. Yang is supported by the Scientific Research Fund Project of Yunnan Education Department (Project ID:\,2025Y0106, KC-24248416).
Tie Liu acknowledges the supports by the PIFI program of Chinese Academy of Sciences through grant No. 2025PG0009, and the Tianchi Talent Program of Xinjiang Uygur Autonomous Region.
GG gratefully acknowledges support by the ANID BASAL project FB210003. 
PS was partially supported by a Grant-in-Aid for Scientific Research (KAKENHI Number JP23H01221) of JSPS.
This work was performed in part at the Jet Propulsion Laboratory, California Institute of Technology, under contract with the National Aeronautics and Space Administration (80NM0018D0004).
SRD acknowledges support from the Fondecyt Postdoctoral fellowship (project code 3220162) and ANID BASAL project FB210003.
This work is sponsored (in part) by the Chinese Academy of Sciences (CAS), through a grant to the CAS South America Center for Astronomy (CASSACA) in Santiago, Chile.
LZ, PG and JH are supported by Chinese Academy of Sciences South America Center for Astronomy (CASSACA) Key Research Project E52H540201 and the China-Chile Joint Research Fund (CCJRF 2211). CCJRF is provided by the CASSACA and established by National Astronomical Observatories, Chinese Academy of Sciences (NAOC) and Chilean Astronomy Society (SOCHIAS) to support China-Chile collaborations in astronomy. 
LB gratefully acknowledges support by the ANID BASAL project FB210003.
This paper makes use of the following ALMA data: ADS/JAO.ALMA\#2019.1.00685.S and ADS/JAO.ALMA\#2021.1.00095.S.

\appendix

Gaussian fitting of the average spectra of $\rm H^{13}CO^+$ molecular emission from the C2 core using the ATOMS data and $\rm N_2D^+$ molecular emission from the condensation using the QUARKS data.

 \begin{figure*}[ht!]
    \centering
    \includegraphics[angle=0, width=0.4\textwidth]{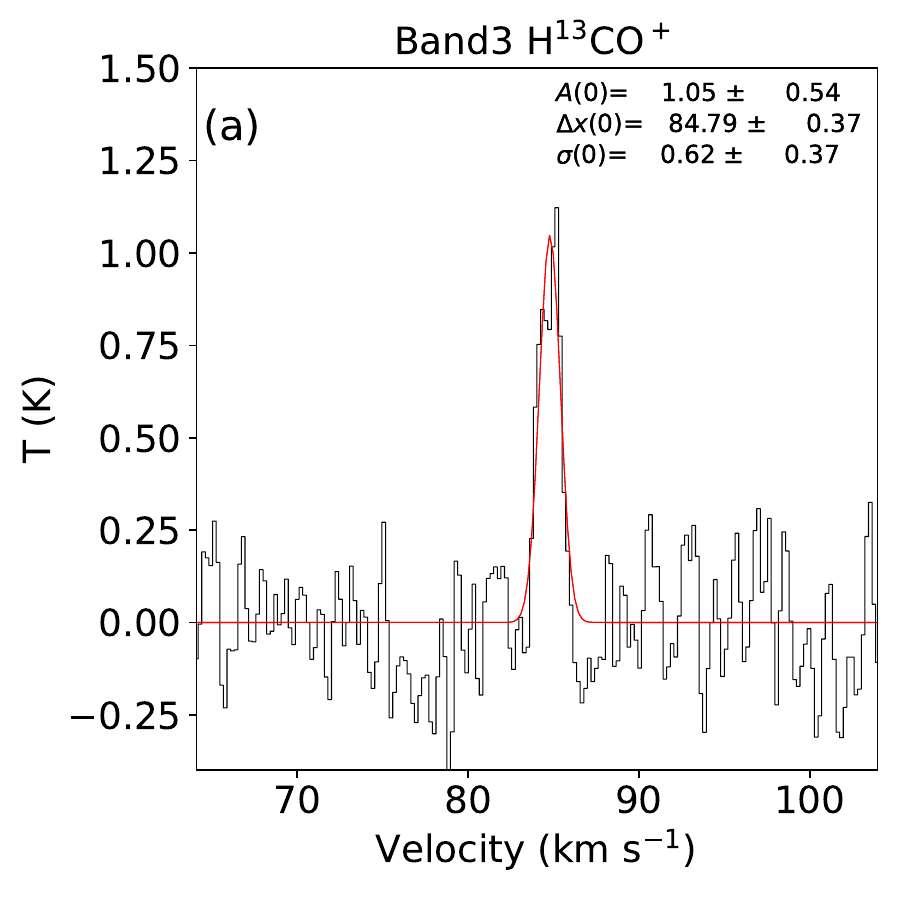} 
    \includegraphics[angle=0, width=0.4\textwidth]{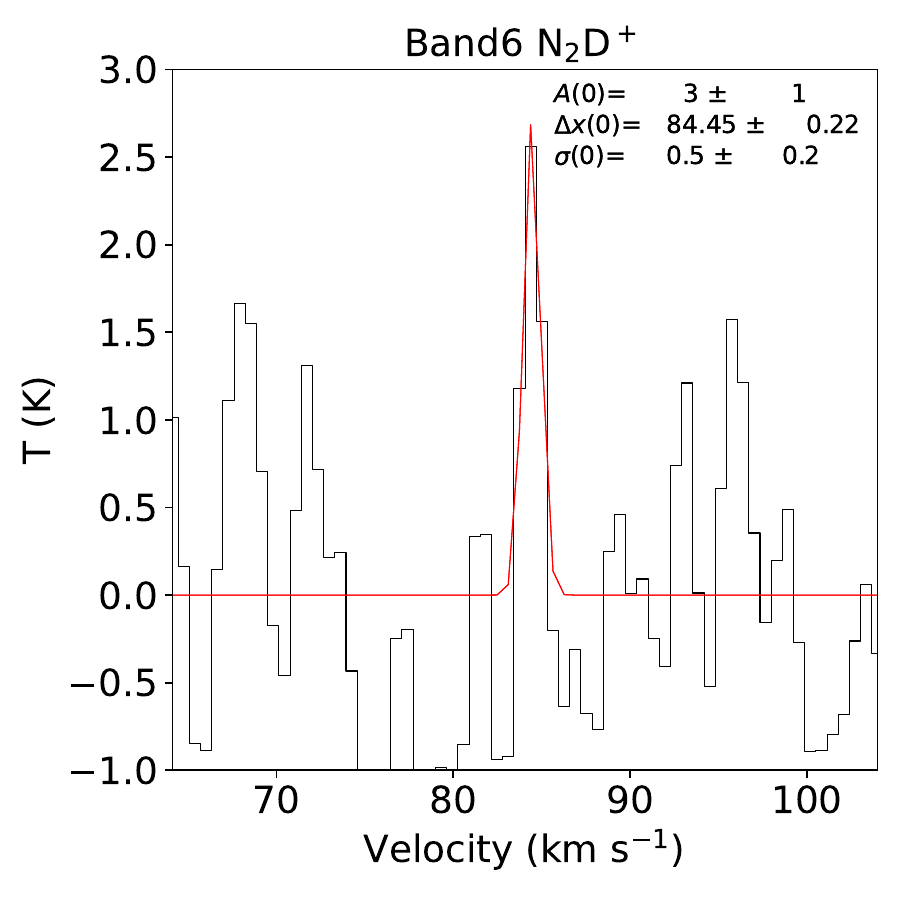} 
    \caption{Average spectra of $\rm H^{13}CO^+$ from the ATOMS survey and average spectra of $\rm N_2D^+$ from the QUARKS survey.} 
    \label{fig:h13co}    
\end{figure*}

\bibliographystyle{aasjournal}
\bibliography{reference}{}

\end{document}